\documentclass[aps,preprint,superscriptaddress,showpacs]{revtex4}
\usepackage{amsmath}
\usepackage{graphicx}
\begin{document}
\title{Twisted Dust Acoustic Waves in Dusty Plasmas}
\author{P. K. Shukla}
\affiliation{International Centre for Advanced Studies in Physical Sciences \& Institute for Theoretical Physics, 
Faculty of Physics and Astronomy, Ruhr University Bochum, D-44780 Bochum, Germany}
\affiliation{Department of Mechanical and Aerospace Engineering \& Centre for Energy Research, 
University of California San Diego, La Jolla, CA 92093, U. S. A.}
\email{profshukla@yahoo.de}
\received{19 July 2012}
\accepted{1 August 2012: Phys. Plasmas {\bf 19}, No. 9, 2012}
\begin{abstract}
We examine linear dust acoustic waves (DAWs) in a dusty plasma with strongly correlated dust grains, and discuss possibility of a  twisted DA vortex beam carrying orbital angular momentum (OAM). For our purposes, we use the Boltzmann distributed electron and ion density perturbations, the dust continuity and generalized viscoelastic dust momentum equations, and Poisson's equation to  obtain a dispersion relation  for the modified DAWs. The effects of the polarization force, strong dust couplings, and dust charge 
fluctuations on the  DAW spectrum are examined. Furthermore, we demonstrate that the DAW can propagate as a twisted vortex beam  carrying OAM. A twisted DA vortex structure  can trap and transport dust particles in dusty plasmas. 
\end{abstract}
\maketitle

\section{Introduction}

Charged dust grains \cite{Langmuir} and dusty plasmas \cite{Shuklamamun} are ubiquteous in cosmic and astrophysical   environments \cite{Alfven,Spitzer78,Evans}, such as interstellar media, molecular dusty clouds, star forming dust clouds,  Eagle nebula, and supernovae remnants, etc. They are also found in our solar system, e.g. in planetary rings  systems \cite{Mendis80,Goertz89,Horanyi04}, in interplanetary media \cite{Mann11} due to the presence of cometary dust  particles, on the Martian surface as dust devils, on the surface of Sun and on moon, as well as in the Earth's   mesosphere \cite{Havnes,Smiley03,Knapp11}, in space as charged dust debris \cite{Juha97} produced when satellites are  destructed,  and near space propulsion vehicles \cite{Ergun10} [for future spacecrafts that go nearer to the Sun  (such as  Solar Orbital and Solar Probe Plus) and Lunar Atmosphere and Dust Environment Explorer mission to the moon]  for exploring  composition of dust grains and their role in collective dust-plasma interactions) due to rocket exhausts. Charged dust particles, which could be of different sizes (ranging from micron-sized to nanometer sized), 
are naturally  formed in industrial processing \cite{Watanabe97} for nanotechnology and in magnetic fusion reactors \cite{Krash11}.  Furthermore, low-temperature dusty plasmas are also produced in laboratory devices \cite{Xu92} for fundamental  studies  in a new environment (e.g. on board International Space Station for examining the behaviour of dusty  plasmas under  microgravity conditions \cite{Fortov04}) that does not exist in the usual electron-ion plasma without dust.  It emerges that dusty plasmas are of broad interdisciplinary interest in physical sciences, and share ideas  with  other fields, e.g. condensed matter physics and astrophysics. Furthermore, dusty plasmas are also finding applications  in medical and biological sciences \cite{Lar99}.
 
A neutral dust particle in plasmas is charged  both negative and positive due to a variety  of physical 
processes \cite{Rosen62,Whipple81,Merlino94,Mendis94,Ost01,Sicka01}, including absorption  of electrons from  the  background plasma \cite{Rosen62}, photo emissions \cite{Knapp11}, triboelectric 
effects \cite{Sicka01}, etc. A dusty plasma is usually composed of electrons, positive ions, negative 
or positive dust grains, and neutral atoms. When  the interaction potential energy ($=Z_d^2 e^2/d$, 
where $Z_d$ is the dust charge state, $e$ the magnitude of the electron charge, and $d$ the inter-grain 
spacing or the Wigner-Seitz radius) between two neighboring dust  particles is much larger (smaller) 
than the dust kinetic energy  $k_B T_d$, where $k_B$ is the Boltzmann constant and $T_d$ the 
dust particle temperature, the dusty plasma is in a strongly (weakly) coupled state. 

More than half a century ago, Wuerker {\it et al.} \cite{Wuerker59} showed that electrically charged iron  and aluminium  particles having diameters  of a few micrometers can be contained in a confined space by alternating and dc electric fields.  Under the three-dimensional focusing alternating gradient focusing force and the Coulomb  repulsive force, charged iron and  aluminium particles form a crystalline array, which can be melted and reformed. This seems to be the first indication of  ordered dust particle structures in an external confined potential. However, following  Ichimaru's idea \cite{Ichimaru82}  of one-component strongly coupled electron system, Ikezi \cite{Ikezi86} postulated the solidification of charged dust particles 
in a dusty  plasma \cite{Shuklamamun,Mendis02,Vladimirov05,Fortov05,Shukla09}, when the dusty plasma 
$\Gamma=Z_d^2e^2/d k_B T_d$ is close to 172. Such values of $\Gamma$ can be achieved in low-temperature laboratory dusty plasma  discharges at room temperatures owing to the large $Z_d$ a micron-size dust grain   would acquire by  absorbing electrons from  the background plasma. The formation of dust Coulomb crystals and ordered dust particle structures have since been observed in  the sheath region of many laboratory  experiments \cite{Chu94a,Chu94b,Thomas94,Hayashi94,Fortov96,Rahman}. The ordered dust  particle structures  are attributed to attractive forces \cite{Nambu95,Shukla96,Shukla09} between negative dust grains due to ion focusing and ion wakefields in a dusty plasma sheath with streaming ions, 
as well as due to overlapping Debye spheres \cite{Resendes98} and dipole-dipole interactions \cite{Shuklamamun}.
 
The collective behavior of dusty plasmas involving an ensembles of charged particles was recognized 
through the prediction of  the dust acoustic wave (DAW) by Shukla \cite{Capri} at the First Capri Workshop  
on Dusty  Plasmas in May of 1989, where he suggested the existence of the nonlinear DAW in the presence 
of  Boltzmann  distributed electrons and ions, and massive charged dust particles. Shukla's idea was then 
worked out in the  first paper \cite{rao90} on linear and nonlinear DAWs. It must be  stressed that there 
does not exist a counterpart of the DAW in an electron-ion plasma without charged dust  grains, 
since the DAW is supported by the dust particle inertia, and the restoring force comes from the  
pressures of the inertialess Boltzmann distributed electrons and ions. Thus, similar to the 
Alfv\'en wave in an electron-ion magnetoplasma without dust, the DAW is of fundamental importance 
in laboratory and space plasmas. The DAW is usually excited by an ion streaming instability \cite{ram92,Rosenberg93},  and has a frequency much smaller than the dust plasma frequency, extending into the infra-sonic frequency range  when the dust particles are anomalously heated. The low-frequency 
(of the order of 10 Hz) DA fluctuations  were first  indirectly observed  in the experiment of  Chu  {\it et al.} \cite{Chu94a} prior to dust particle  crystallization,  and have since been spectacularly observed in many laboratory experiments world-wide \cite{Barkan,Fortov,Prab},  and also in the Earth's ionosphere \cite{Popel09}. Thus, the existence of the DAW in dusty plasmas have been  demonstrated at kinetic levels, and the visual images of the dust acoustic wavefronts by naked eyes are  possible \cite{Barkan}. 

In this paper, we revisit the linear DAW in a dusty plasma by incorporating the effects of dust particle 
correlations \cite{Kaw98}, the polarization force \cite{Khrapak} due to  interactions between thermal ions 
and highly charged dust grains, and dust charge fluctuations (DCFs) \cite{Varma93}.   A linear dispersion relation is derived and analyzed. The underlying physics of the DAW has been put on the firm footing.  Furthermore, we also discuss the possibility of  a twisted dust acoustic wave (TDAW) carrying OAM [we also refer it   as a dust acoustic vortex (DAV) beam]. A TDAW or a DAV beam can, in turn, be used for trapping 
and transporting  charged dust grains from one region to another in laboratory and space dusty plasmas.

\section{Theoretical consideration}

Let us consider an unmagnetized dusty plasma composed of inertialess electrons and ions, as well as strongly   correlated negative dust particles of uniform sizes. In the presence of ultra-low frequency DAWs, 
with $ \omega \ll \nu_{en}, \nu_{in} \ll k^2V_{Te, Ti}^2/|\omega|$, where $\omega$ is the wave frequency, 
$\nu_{en}$ $(\nu_{in})$ the electron (ion)-neutral collision frequency, $k$ the wave number, 
and $V_{Te}$ $(V_{Ti})$ the electron  (ion) thermal speed, both electrons and ions obey the Boltzmann 
law (deduced from the balance of the electric force and pressure gradients of the electrons and ions), 
since they can be considered inertialess on the timescale of the DAW period, and rapidly thermalize 
due to collisions with neutrals. Thus, the electron and ion number density perturbations 
($n_{e1,i1} \ll n_{e0,i0}$) are, respectively, given by \cite{rao90}
\begin{equation}
n_{e1}  \approx n_{e0} \frac{e\phi}{k_B T_e},
\end{equation}
and 
\begin{equation}
n_{i1} \approx - n_{i0} \frac{e\phi}{k_B T_i},
\end{equation}
where $n_{e0}$ and $n_{i0}$ are the unperturbed electron and ion number densities, respectively, $\phi$ 
the electrostatic   potential of the DAW, and $T_e$ $(T_i)$ the electron (ion) temperature. At equilibrium, we have the quasi-neutrality  condition \cite{rao90,silin}, viz.  $n_{i0} =n_{e0} + Z_d n_{d0}$,  where $Z_{d0}$ is the average number of electrons  residing on a dust grain, and $n_{d0}$ the  unperturbed dust number density.

The dynamics of dust particles in a our dusty plasma is governed by the hydrodynamic equations composed of 
the  dust continuity equation
\begin{equation}
\frac{\partial n_{d1}}{\partial t} + n_{d0} \nabla \cdot {\bf v}_d =0,
\end{equation} 
and the generalized dust fluid momentum equation 
\begin{eqnarray}
 \left(1+ \tau_m \frac{\partial}{\partial t} \right) \left[\frac{\partial {\bf v}_d } {\partial t} 
+ \nu_d {\bf v}_d  - \frac{Z_{d0} e}{m_d}\ (1-R) \nabla \phi 
 + \frac{\mu_d k_B T_d}{\rho_d}\nabla  n_{d1}  \right]\nonumber \\
= \frac{\eta}{\rho_d} \nabla^2 {\bf v}_d + \frac{\left(\xi + \frac{\eta}{3}\right)}{\rho_d}
 \nabla (\nabla \cdot {\bf v}_d),
\end{eqnarray}
where $n_{d1} (\ll n_{d0})$ and ${\bf v}_d$ are the dust number density and dust fluid velocity perturbations, 
respectively, $m_d$ the dust mass, $\rho_d =n_{d0} m_d$ the dust mass density, $R =Z_{d0} e^2/4 k_B 
T_i \lambda_{Di}$ is a parameter determining the effect of the polarization force \cite{Khrapak} that arises 
due to the interaction between thermal ions and negative dust grains, $\mu_d n_{d0} k_B T_d\equiv P_d$ 
the  effective dust thermal pressure, $\mu_d =1+(1/3) u (\Gamma) +(\Gamma/9)\partial u(\Gamma)/\partial \Gamma$  the compressibility, $\Gamma=Z_{d0}^2 e^2/d k_B T_d$ the ratio between the dust Coulomb 
and dust thermal energies,  $d = (3/4\pi n_{d0})^{1/3}$ the Wigner-Seitz radius, $u(\Gamma)$ is a 
measure  of the excess internal energy of the system, which reads \cite{Abe,Slatt} $u(\Gamma)\simeq -(\sqrt{3}/2)\Gamma^{3/2}$   for $\Gamma \leq 1$ ({\it viz}. a liquid-like state), and $u(\Gamma) 
=-0.80 \Gamma + 0.95 \Gamma^{1/4}  + 0.19 \Gamma^{-1/4} -0.81$ in a range $1 < \Gamma < 200$. 
The coupling parameter in dusty plasmas including the shielding of a negative dust grain by electrons and ions reads $\Gamma_g = \Gamma \exp(-\kappa)$, where $\kappa = a_d/\lambda_D$, $a_d$ is the inter-dust grain spacing, and $\lambda_D= \lambda_{De}\lambda_{Di}/(\lambda_{De}^2+\lambda_{Di}^2)^{1/2}$ the effective Debye radius of dusty plasmas, with $\lambda_{De}=(k_B T_e/4 \pi n_{e0} e^2)^{1/2}$ and $\lambda_{Di}= (k_B T_i/4 \pi n_{i0} e^2)^{1/2}$ being the ion  and electron Debye radii, respectively. 
The dust-neutral collision frequency  is \cite{Baines65}  $\nu_{dn}=(8/3)\sqrt{2\pi} m_n n_n r_d^2 V_{Tn}/m_d$, where  $m_n$ is the neutral mass, $n_n$ the neutral number density,  $r_d$ the dust grain radius,  $V_{Tn}=(k_B T_n/m_n)^{1/2}$ the neutral thermal speed, and $T_n$ the neutral gas temperature.  The visco-elastic properties of the dust fluids are characterized by the relaxation time \cite{Ichimaru86,Berk}  $\tau_m = [ (\xi +4\eta/3)/n_{d0}T_d]/\left[1-\mu_d + 4 u(\Gamma)/15\right]$, involving the shear and bulk viscosities $\eta$ and $\xi$. There are various approaches for calculating $\eta$ and $\xi$, which are widely discussed in  the literature \cite{Slatt}. We note that the generalized viscoelastic dust momentum Eq. (4) is an extension  of Kaw and Sen \cite{Kaw98}, by including the effects of the polarization force \cite{Khrapak}. The viscoelastic momentum equation, similar to Eq. (4), has also been used in the study of collective phenomena in fluids \cite{Frenkel} and in one-component plasmas with  strongly correlated electrons \cite{Ichimaru86,Berk}.

The DA wave potential $\phi$ is obtained  from Poisson's equation
\begin{equation}
\nabla^2 \phi = 4 \pi e \left(n_{e1} - n_{i1} + Z_{d0} n_{d1} + Z_{d1} n_{d0} \right),
\end{equation}
where the dust charge perturbation $Z_{d1}  (\ll Z_{d0})$ is determined from the charging equation \cite{Shuklamamun}
\begin{equation}
\left(\frac{\partial}{\partial t}  +\nu_1 \right) Z_{d1} = \frac{a_d \nu_2}{e} \phi.
\end{equation}
Here we have introduced the notations
\begin{equation}
\nu_1 = a_d \left[\frac{\omega_{pe}}{\lambda_{De}} \exp(-\eta_e) + \frac{\omega_{pi}}{\lambda_{Di}}\right],
\end{equation}
and
\begin{equation}
\nu_2 = a_d \left[\frac{\omega_{pe}}{\lambda_{De}} \exp(-\eta_e) + (1+ \eta_i) 
\frac{\omega_{pi}}{\lambda_{Di}}\right],
\end{equation}
with $\eta_{e,i} =Z_{d0}e^2/a_d k_B T_{e,i}$. Equation (6) reveals that the DCFs cause an adverse phase 
lag between $Z_{d1}$ and $\phi$, which lead to the DAW damping \cite{Varma93}. However, since the DAW frequency is much smaller  that  the dust charging frequency $\nu_1$, one notices that $Z_{d1}$ and $\phi$ are in phase. Subsequently, there appears a decrease of the DAW phase speed, as shown in the  subsection below,  

From Eqs. (1), (2), (5) and (6) we readily obtain
\begin{equation}
\left[(\frac{\partial}{\partial t} + \nu_1)(\nabla^2 - k_D^2) - k_q^2 \nu_2 \right] \phi
= 4\pi e Z_{d0}(\frac{\partial}{\partial t} + \nu_1) n_{d1},
\end{equation}
where  $k_D =(k_e^2 + k_i^2)^{1/2} \equiv 1/\lambda_D$, $k_{e,i} =1/\lambda_{De,Di}$, a
nd $k_q =(4\pi a_d n_{d0})^{1/2}$.

Furthermore, from Eqs. (3) and (4) we have
\begin{eqnarray}
 \left(1+ \tau_m \frac{\partial}{\partial t} \right) \left[ (\frac{\partial^2} {\partial t^2}
- V_{Td}^2 \nabla^2) n_{d1} + \frac{Z_{d0} e (1-R) }{m_d}\nabla^2 \phi \right]\nonumber \\
= - \left[\nu_d - \frac{(\xi + 4\eta/3)}{\rho_d} \nabla^2\right] \frac{\partial n_{d1}}{\partial t} ,
\end{eqnarray}
where $V_{Td} = (\mu_d k_B T_d/m_d)^{1/2}$ is the effective dust thermal speed. 
Equations (9) and (10) are  the desired equations governing the linear propagation of the modified DAWs. 

\subsection{Modified DAWs}

Within the framework of a plane-wave approximation, assuming that $n_{d1}$ and $\phi$ are 
proportional to  $\exp(-i\omega t + i {\bf k} \cdot {\bf r})$, we Fourier analyze (9) and (10) and 
combine the resultant equations to obtain the dispersion relation for the modified DAWs
\begin{equation}
1 +k^2 \lambda_D^2 + \frac{k_q^2 \lambda_D^2\nu_2}{(\nu_1-i \omega)}
- \frac{\omega_{d}^2}{[\omega(\omega + i \nu_d)-k^2 V_{Td}^2 +  i \omega \omega_v]} =0,
\end{equation}
where $\omega_d^2 = k^2 C_d^2(1-R) > 0$, $C_d =\omega_{pd}\lambda_D$, 
$\omega_v =\eta_*k^2/(1-i \omega \tau_m)$, and $\eta_* =(\xi +4\eta/3)/m_d n_{d0}$. 

Several comments are in order. First, for $\nu_d, \ll |\omega| \ll 1/\tau_m, \nu_1$, we have from (11)
\begin{equation}
\omega^2 = k^2 U_d^2 + \frac{\omega_d^2}{1 + (k^2 +k_q^2\nu_2/\nu_1)\lambda_D^2},
\end{equation}
where $U_d^2 = V_{Td}^2 +\eta_*k^2/\tau_m$. Second, in the long wavelength limit, viz.
$k^2 \lambda_D^2 \ll 1$, Eq. (12) reduces to
\begin{equation}
\omega^2 = k^2 U_d^2 + \frac{\omega_d^2}{(1 + k_q^2 \lambda_D^2 \nu_2/\nu_1)}.
\end{equation}
Third, the well-known frequency of the DAW, in the absence of dust grain correlations,
dust fluid viscosities, dust-neutral collisions, polarization and ion pressure effects, 
and DCFs, can be obtained from Eq. (11). We have the famous result \cite{rao90}

\begin{equation}
\omega = \frac{k C_d}{(1+k^2 \lambda_D^2)^{1/2}}.
\end{equation}
Since $C_d = Z_{d0} \left[n_{d0}k_B T_i /n_{i0} m_d (1+\alpha_1)\right]^{1/2}$, where
$\alpha_1=n_{e0}T_i/n_{i0}T_e$, we observe from Eq. (14) that the phase speed ($\omega/k)$ of 
the long wavelength  (in comparison with $\lambda_D)$ DAWs is proportional to $[k_B T_i/m_d (1+\alpha_1)]^{1/2}$, dictating  that the restoring force in the DAW comes from  the pressures of the inertialess Boltzmann distributed  electron and ion fluids, while the dust mass provides  the inertia 
to sustain the wave. The wave  dispersion  [the $k^2 \lambda_D^2$-term in Eq. (14)] arises owing to the departure from the quasi-neutrality condition  in the perturbed density perturbations.

\subsection{Twisted DAWs}

We now discuss the possibility of a twisted DAV  beam in the long-wavelength  limit 
(viz. $k^2 \lambda_D^2 \ll 1$), in which case the quasi-neutrality holds. Here we have
\begin{equation}
n_{d1} =-\frac{n_{i0}}{Z_{d0} k_B T_i}(1+\alpha_1 + \alpha_2)\phi,
\end{equation}
where $\alpha_2 = a_d \nu_2 k_B T_i n_{d0}/\nu_1 e^2 n_{i0}$. Equation (15) depicts that the dust density 
compression is possible,  since $\phi < 0$ (due to negative  charge on dust grains) for the DAWs. 

In order to study the property of a TDAW, we consider the limit $\lambda_D^2 \nabla^2 \phi \ll \phi$
 in Eq. (9) and combine the resultant equation with Eq. (10) to obtain the simple wave equation
\begin{equation}
\left(\nabla^2 + K^2 \right)n_{d1} =0,
\end{equation}
where we have assumed that $n_{d1}$ is proportional to $\exp(-i \omega t)$, and have denoted 
$K^2 =\omega^2/U_*^2$ with $U_*^2=V_{Td}^2 +C_d^2(1-R)/(1+k_q^2\lambda_D^2\nu_2/\nu_1)$.
We have assumed that $k^2 \lambda_D^2 \ll 1$ and $\nu_d, \eta_* k^2  \ll |\omega| \ll 1/\tau_m,\nu_1 $. 

We now seek a solution of Eq. (16) in the form
\begin{equation}
n_{d1} = n_l (r)\exp(i k z),
\end{equation}
where $n_l (r)$ is a slowly varying function of $z$. Here $r =(x^2 + y^2)^{1/2}$ and $k$  is the 
propagation wave number along the axial ($z-$) direction. By using Eq. (17) one can write Eq. (16) 
in a paraxial approximation as
\begin{equation}
\left(2i k \frac{\partial}{\partial z} +\nabla_\perp^2\right)  n_l = 0,
\end{equation}
where $k =K \equiv \omega/U_*$ and $\nabla_\perp^2 =(1/r) (\partial/\partial r) (r\partial/\partial r)
+ (1/r^2)\partial^2/\partial \theta^2$. We have used cylindrical coordinates with ${\bf r} =(r,\theta,z)$.

The solution of Eq. (18) can  be written as a superposition of Laguerre-Gaussian (LG) modes \cite{Allen92}, 
each of them representing a state of orbital angular momentum, characterized by the quantum number $l$, 
such that

\begin{equation}
n_l  =\sum_{pl} n_{pl} F_{pl}(r,z) \exp(il \theta),
\end{equation}
where the mode structure function is

\begin{equation}
F_{pl}(r,z) = F_{pl} X^{|l|} L_p^{|l|}(X) \exp(-X/2),
\end{equation}
with $X=r^2/w^2(z)$, and $w(z)$ is the DA beam width. The normalization factor $ F_{pl}$
and the associated Laguerre polynomial $L_p^{|l|}(x)$ are, respectively,

\begin{equation}
F_{pl} = \frac{1}{2\sqrt{\pi}}\left[\frac{(l+p)!}{p!}\right]^{1/2},
\end{equation}
and

\begin{equation}
L_p^{|l|}(X) = \frac{\exp(X)}{X p!} \frac{d^p}{dX^p}\left[X^{l+p} \exp(-X)\right],
\end{equation}
where $p$ and $l$ are the radial and angular mode numbers of the DAW orbital angular momentum state.
In a special case with $l=0$ and $p=0$, we have a Gaussian beam.

The LG solutions, given by Eq. (19), describe the feature of a twisted DAV beam carrying OAM. 
In a twisted DAV beam, the wavefront would rotate around the beam's propagation direction in a spiral that looks like  fusilli pasta (or a bit like a DNA double helix), creating a vortex and leading to the DAV beam with zero intensity at  its center. A twisted DAV beam can be created with the help of two oppositely propagating three-dimensional DAWs   that are colliding in a dusty plasma. Twisting of the DAWs would occur because different sections of the wavefront  would bounce off different steps, introducing a delay between the reflection of neighboring sections and, therefore,  causing the wavefront to be twisted due to an entanglement of the wavefronts, and take on the shape of the reflector.  Thus, due to angular symmetry, Noether  theorem guards OAM conservation even for a longitudinal DAV beam.

The energy flux of the DAV beam is given by $ W {\bf V}_g $, where ${\bf V}_g (=\partial \omega/\partial {\bf k}$)  is the group velocity of the DAV beam and its energy density reads \cite{Mendonca}
\begin{equation}
 W = \frac{\partial}{\partial \omega}\left[\omega \epsilon (\omega, {\bf k})\right]_{\omega = \omega_k}
\frac{|{\bf E}|^2}{4\pi},
\end{equation}
where ${\bf E} =  i {\bf k} \phi$, and for $\nu_d, \ll |\omega| \ll 1/\tau_m, \tau_1$ and $k^2 \lambda_D^2 \ll 1$ we have  $\epsilon (\omega, {\bf k}) = (1/k^2\lambda_D^2) +(k_q^2\nu_2/k^2\nu_1)-\omega_{pd}^2(1-R)/(\omega^2 -k^2 V_{Td}^2)$. 
Equation (15) exhibits a relationship between the electrostatic potential$\phi$ and the dust density perturbation $n_{d1}$.  It turns out that the energy flux is independent of the mode number $l$.

\section{Summary and conclusions}

To summarize, we have presented a new dispersion relation (11) for the DAWs in an unmagnetized dusty plasma that is  composed of weakly  correlated  Boltzmann distributed electrons and ions in the wave potential, and strongly correlated highly-charged dust grains  that follow the viscoelastic dust momentum equation. In our investigation,  the effects of the polarization force and DCFs  are incorporated. It is found that the contributions of both the polarization force and DCFs are to reduce the  frequency of the DAWs. Furthermore, we have discussed the possibility of a twisted DAV beam. The latter can  trap charged dust 
particles and transport them from one region to another. The present study of a twisted DAV beam can be useful   for diagnostic purposes when the DAW frequencies are near the infra-sonic frequencies in laboratory, space and  cosmic dusty plasmas.  In closing, we mention that the importance of OAM of electromagnetic waves in the astrophysical context was recognized by  Harwit \cite{Harwit}, and is also the subject of current interest \cite{Leckner,An} in connection with twisted ultrasound pulses.

\acknowledgments
The author thanks Robert Merlino, Jose Tito Mendon\c{c}a, and Lennart Stenflo for useful discussions. This research was partially 
supported by the Deutsche Forschungsgemeinschaft (DFG), Bonn, through the  project SH21/3-2 of the  Research Unit 1048.

\end{document}